\documentclass[11pt]{article}

\usepackage[paper=a4paper,left=20mm,right=20mm,top=25mm,bottom=25mm]{geometry}
\usepackage{amsmath} 
\usepackage{amsthm}
\usepackage{amssymb}

\usepackage{graphicx}
\usepackage{color}
\usepackage{eufrak}
\numberwithin{equation}{section}

\def\ben{\begin{enumerate}} \def\een{\end{enumerate}}
\def\beq{\begin{equation}} \def\eeq{\end{equation}}
\def\beqn{\begin{equation*}} \def\eeqn{\end{equation*}}
\def\bea{\begin{eqnarray}} \def\eea{\end{eqnarray}}
\def\ba{\begin{array}} \def\ea{\end{array}}
\def\beann{\begin{eqnarray*}} \def\eeann{\end{eqnarray*}}
\def\beasn{\begin{sneqnarray}} \def\eeasn{\end{sneqnarray}}

\def\bi{\begin{itemize}} \def\ei{\end{itemize}} 
\def\bd{\begin{description}} \def\ed{\end{description}}
\def\be{\begin{enumerate}} \def\ee{\end{enumerate}}

\def\ea{\'e}

\def\mf{\mathfrak}

\def\ea{\'e}

\def\buildchar#1#2#3{\null \! \mathop {\vphantom {#1}\smash
#1}\limits ^{#2}_{#3}\!\null }
\def\OT#1{\buildchar{{#1}}{\;_\sim}{}\/}
\def\UT#1{\buildchar{{#1}}{}{^\sim}\/}
\def\OTT#1{\buildchar{{#1}}{\;_\approx}{}\/}

\usepackage{graphicx}
\usepackage{color}

\title{Diffeomorphism covariance of the canonical Barbero-Immirzi-Holst triad theory}

\author
{Donald Salisbury\\
\\
\normalsize{Austin College, 900 North Grand Ave, Sherman, Texas 75090, USA}\\
}

\begin{document}

\maketitle

\begin{abstract}
The vanishing phase space generator of the full four-dimensional diffeomorphism-related symmetry group in the context of the Barbero-Immirz-Holst Lagrangian is derived directly for the first time from Noether's second theorem. It's applicability in the construction of classical diffeomorphism invariants is reviewed.

\end{abstract}

\section{Introduction}

What I identify as the Barbero-Immirzi-Holst model serves as a foundation for today's canonical approach to loop quantum gravity. I will derive in this article a new analysis of the underlying four-dimensional spacetime diffeomorphism-related classical canonical symmetry. I will derive the canonical symmetry generators directly from the vanishing charge that follows from Emmy Noether's second theorem, in a manner similar to the first such derivation presented for conventional canonical gravity in \cite{Salisbury:2022aa}. The focus will be on a reformulated ADM approach that incorporates densitied triads. And I will argue that the extension of this analysis to the new triad approach to gravity as proposed by \cite{Barbero:1995aa}, \cite{Immirzi:1997aa}, and \cite{Holst:1996aa} is almost trivial. As is well known, in order to achieve the results of canonically generated variations of spacetime coordinates it is necessary to supplement the variations of phase space variables under diffeomorphims with related triad gauge transformations. I conclude with an overview of a technique for introducing intrinsic coordinates as gauge conditions, and employing the full diffeomorphism generator to construct invariant temporal evolution in a manner related to Rovelli's relative observables \cite{Rovelli:2002ab}. This lays the foundations for an eventual application in loop quantum gravity.

  \section{Derivation of canonical Hamiltonian}
  
 I use the ADM Lagrangian as rewritten using triad variables.
\beq
{\cal L}_{ADM} = N t \left({}^3\!R + K_{ab} K^{ab} - \left(K^a_a\right)^2\right)
= N t \left({}^3\!R + K_{ab} e^{ac}e^{bd}K_{cd} - \left(e^{ab}K_{ab}\right)^2\right),
\eeq
where
\beq
K_{ab} = \frac{1}{2N} \left( g_{ab,0} - N^c g_{ab,c} - g_{ca} N^c_{,b} -g_{cb} N^c_{,a} \right)
= \frac{1}{2N} \left(g_{ab,0} -  2g_{c(a} N^c_{|b)} \right). 
\eeq
The variable $t$ is the determinant of the spatial metic $g_{ab}$, with $e^{ab}$ its inverse. The variable $N$ is the lapse while $N^a$ represents the metric shift functions. ${}^3\!R $ is the tree-dimensional curvature scalar.
 
The first task is to specialize to tetrads with the choice $E^\mu_0 = n^\mu = \delta^\mu_0 N^{-1} - \delta^\mu_a N^{-1} N^a$. This tetrad is orthogonal to the constant time hypersurface. The covariant metric is
\beq
    g_{\mu\nu} = \begin{pmatrix}
             -N^2 + N^{c}N^{d}g_{cd} & g_{ac}N^{c} \\
                              g_{bd}N^{d}  & g_{ab}  \end{pmatrix},
\eeq
with the contravariant metric
\beq
 g^{\mu\nu} =  \begin{pmatrix}-1/N^2 & N^a/N^2 \\
                              N^b/N^2 & e^{ab} - N^aN^b/N^2 \end{pmatrix}.
\eeq
We then choose the remaining tetrads to be tangential to the constant time hypersurface. Thus the full set of contravariant tetrads (with the upper index representing the row and the lower index representing the column) is
\beq
E^\mu_I =  \begin{pmatrix}
N^{-1} & 0 \\
-N^{-1} N^a & T^a_i
\end{pmatrix}
\eeq 
with the corresponding covariant set
\beq
e^I_\mu =  \begin{pmatrix}
N & 0 \\
t^i_a N^a & t^i_a
\end{pmatrix}
\eeq

We shall, however, employ as independent triad variables $\OT T^a_i :=  t T^a_i$ where $t := \det \left( t^i_a \right)$. Furthermore, rather than choosing the lapse $N$ as an independent configuration variable we work with $\UT N := t^{-1}N$. So for the following we will need
\beq
t_{,\mu} = t t^i_{a,\mu} T^a_i =  \left(t^i_{a} \tilde T^a_i\right)_{,\mu} - t^i_{a} \tilde T^a_{i,\mu},
\eeq
so we find that
\beq
t_{,\mu} = \frac{1}{2} t^i_a \tilde T^a_{i,\mu},
\eeq
\beq
t^i_{a,\mu} = 2 t^{-1} t^{[i}_a t^{j]}_b \tilde T^b_{j,\mu},
\eeq
and
\beq
T^a_{i,\mu} = -\frac{1}{2} t^{-2} t^j_b \tilde T^b_{j,\mu} \tilde T^a_i + t^{-1} \tilde T^a_{i,\mu}.
\eeq
 Now define the canonical momentum
\bea
p^l_e &:=& \frac{\partial {\cal L}_{ADM}}{\partial  \tilde T^e_{l,0}} \nonumber \\
&=& 2Nt \left(e^{ac}e^{bd} - e^{ab} e^{cd}\right) K_{cd}\frac{\partial K_{ab} }{\partial  \tilde T^e_{l,0}} \nonumber. \\
\eea
So I need
\beq
2Nt\frac{\partial K_{ab} }{\partial  \tilde T^e_{l,0}} =  g_{ab}  t^l_e - 2 t^l_{(a} g_{b)e}.
\eeq
 Therefore
 \beq
 p^l_e = - 2 T^d_lK_{ed}, 
 \eeq
 from which we deduce that
 \beq
 p^i_a t^i_b = - 2 K_{ab}.
 \eeq
 So I can write the Lagrangian immediately in terms of the canonical momenta. 
 
To obtain the canonical Hamiltonian ${\cal H}_c$ I must now focus on $p^i_a \tilde T^a_{i,0}$ which I want to write in terms of the momenta. I have
 \beq
 p^i_a \tilde T^a_{i,0} = -2 K_{ab} T^b_i \tilde T^a_{i,0}
 \eeq
 I will rewrite this in terms of derivatives of $t^j_c$. So consider first
 \bea
 \tilde T^a_{i,0} = \left(t T^a_i \right)_{,0} = t_{,0} T^a_i + t T^a_{i,0} = t t^j_{c,0} T^c_j T^a_i - t T^c_i T^a_j t^j_{c,0},
 \eea
 and I therefore have
 \beq
 p^i_a \tilde T^a_{i,0} =- t K_{ab}\left(e^{ab} e^{cd} - e^{bc} e^{ad}\right)g_{cd,0}
 \eeq
 But
 \beq
 g_{cd,0} = 2 N K_{cd} + 2 g_{e(c} N^e_{|d},
 \eeq
 so I conclude finally that
 \beq
 p^i_a \tilde T^a_{i,0} = -2 t K_{ab}\left(e^{ab} e^{cd} - e^{bc} e^{ad}\right) \left(N K_{cd} +  g_{e(c} N^e_{|d}\right)
 \eeq
 I thereby obtain the expression for the canonical Hamiltonian
 \bea
 {\cal H}_c &=& p^i_a \tilde T^a_{i,0}  - {\cal L}_{ADM} \nonumber \\
 &=&  \UT N  \left(-{}^3\!R + K_{ab} e^{ac}e^{bd}K_{cd} - \left(e^{ab}K_{ab}\right)^2\right)  + 2 t\left(-K^a_a e^{cd}+ K^{cd}\right)  g_{ec} N^e_{|d}
 \eea
 For later use I need to rewrite the canonical Hamiltonian in terms of $p^i_a$ using $K_{ab} = -\frac{1}{2} p^i_a t^i_b$, which, implies that
 \beq
 K_{ab} e^{ac}e^{bd}K_{cd} = \frac{1}{4} p^i_a t^i_b p^j_c t^j_d e^{ac}e^{bd} = p^i_a p^i_b e^{ab},
 \eeq
 and
 \beq
 e^{ab} K_{ab} e^{cd} K_{cd} = \frac{1}{4} p^i_a T^a_i p^j_b T^b_j.
 \eeq
 So the canonical Hamitonian becomes
 \beq
  {\cal H}_c=  \UT N  \left(-{}^3\!R + \frac{1}{4}p^i_a p^i_b e^{ab} - \frac{1}{4}p^i_a T^a_i p^j_b T^b_j\right)  +\frac{1}{2} \left(p^i_a T^a_i  e^{cd}- p^i_a T^d_i e^{ac}\right) t g_{e(c} N^e_{|d)}  
 \eeq
 
 (It is straightforward to check that this does deliver an almost correct expression for the time rate of change of the densitized triad - lacking, as we shall see shortly, the arbitrary triad gauge rotations),
 \bea
 \tilde T^e_{l,0} = \frac{\partial  {\cal H}_c}{\partial p^l_e} &=& -2 \UT N\left(e^{ac}e^{bd} - e^{ab} e^{cd}\right) K_{cd}\frac{1}{2} \delta^e_a t^l_b + t g_{f(c} N^f_{|d)}\left(e^{ab} e^{cd} - e^{ac} e^{bd}\right)\frac{1}{2} \delta^e_a t^l_b \nonumber \\
  &=&  - \UT N\left(e^{ec}T^d_l - e^{eb} e^{cd}\right) K_{cd}  +\frac{1}{2} t g_{f(c} N^f_{|d)}\left(T^e_l e^{cd} - e^{ec} T^d_l\right) 
 \eea
 
 It is important to recognize here that the ADM Lagrangian does not depend on the antisymmetrized linear combination of velocities
 $\tilde T^{a[i} \UT t^{j]}_a$,  and as a consequence we will obtain a corresponding primary constraint, with a corresponding addition to the Hamiltonian generator of time evolution.  Rosenfeld had indeed in \cite{Rosenfeld:1930aa} \cite{Rosenfeld:2017ab} considered a tetrad version of general relativity in which analogous constraints appeared and, although he did not explicitly construct the corresponding extended Hamiltonian, it was shown in \cite{Salisbury:2017aa} that he could easily have applied his new techniques to do so. I will next derive the relevant primary constraint by applying Noether's second theorem.
 
 \section{Noether charges}
 
 First there is a vanishing charge that arises from the invariance of the ADM action under triad rotations
\beq
\delta_\eta  T^a_i = \epsilon^{ijk} \tilde T^a_j \eta_k,
\eeq
where the $\eta_k$ are arbitrary spacetime functions.
Following Noether's second theorem, conserved charge arises as follows. The variation of the action is
\beq
0 = \delta_\eta  \int d^4\!x {\cal L}_{ADM} = \int d^4\!x \left[\left(\frac{\delta {\cal L}_{ADM}}{\delta  \tilde T^a_i} \right)  \delta_\eta \tilde T^a_i+ \left(\frac{\partial {\cal L}_{ADM}}{\partial \tilde T^a_{j,\mu}}  \epsilon^{ijk} \tilde T^a_j \eta_k  \right)_{,\mu}\right]
\eeq
When the field equations are satisfied we thus obtain, letting the variations vanish at spatial infinity, the conserved charge
\beq
C_\eta = \int d^3x p^i_a  \epsilon^{ijk} \tilde T^a_j \eta_k.
\eeq
But since $\eta_k$ can vary arbitrarily with time we deduce the existence of constraints
\beq
0 = {\cal H}^k :=  \epsilon^{ijk}p^i_a  \tilde T^a_j .
\eeq

The additional constraints that arise from the invariance of the action under spacetime diffeomorphisms will require a bit more work to derive. I will derive the vanishing Noether charge diffeomorphism-related generator following the procedure that was applied in  the conventional metric case in \cite{Salisbury:2022aa}.  It should be noted here that this procedure was applied to tetrad-based general relativity by Rosenfeld in 1930. And as observed in  \cite{Salisbury:2017aa} he did not complete the derivation of the canonical generators that I will shortly find, very likely because he recognized that he could not express them exclusively in terms of canonical variables. In other words he did not recognize, as first observed in \cite{Pons:1997aa}, that the variations were not projectable under the Legendre transformation to phase space. 

Under an infinitesimal diffeomorphism $x'^\mu = x^\mu - \epsilon^\mu$, the scalar density ${\cal L}_{ADM} $ transforms as\footnote{A major advantage in employing the ADM Lagrangian is that it does vary as a Lagrangian density, assuming only that variations at spatial infinity vanish. See \cite{Kiefer:2012aa}, p. 119 and \cite{Danieli:2020aa} }
\beq
\bar \delta {\cal L}_{ADM} = \left({\cal L}_{ADM} \epsilon^\mu\right)_{,\mu},
\eeq
where the $\bar \delta$ variation is actually the Lie derivative ${\cal L}_\epsilon$. I will shortly work out the corresponding field variations. But first I will derive the corresponding vanishing Noether charges noting that when the field equations are satisfied, and letting $\epsilon^a \rightarrow 0$ at spatial infinity,
\bea
\int d^4\!x \bar \delta {\cal L}_{ADM} &=& \int d^3\!x  \left. \left(\frac{\partial {\cal L}_{ADM}}{\partial \tilde T^a_{i,0}}  \bar \delta \tilde T^a_i  +\frac{\partial {\cal L}_{ADM}}{\partial \UT N_{,0}}  \bar \delta \UT N +\frac{\partial {\cal L}_{ADM}}{\partial  N^a_{,0}}  \bar \delta N^a\right)  \right|^{x^0_f}_{x^0_i}\nonumber \\
&=&  \int d^3\!x  \left. {\cal L}_{ADM} \epsilon^0  \right|^{x^0_f}_{x^0_i}
\eea
So again taking into account that the time dependence of $\epsilon^\mu$ is arbitrary we derive the corresponding vanishing Noether charges
\beq
C_\epsilon =  \int d^3\!x \mf{C}_\epsilon
\eeq
with vanishing charge density
 \bea
\mf{C}_\epsilon &=& \frac{\partial {\cal L}_{ADM} }{\partial \tilde T^a_{i,0}} \bar \delta  \tilde T^a_i+ \frac{\partial {\cal L}_{ADM} }{\partial \UT N_{,0}} \bar \delta \UT N+ \frac{\partial {\cal L}_{ADM} }{\partial  N^a_{,0}} \bar \delta N^a- {\cal L}_{ADM} \epsilon^0 \nonumber \\
&=& p^i_a   \bar \delta  \tilde T^a_i + \OTT P \bar \delta \UT N + \OT P_a \bar \delta N^a - {\cal L}_{ADM} \epsilon^0  
\eea
We recognize, of course, that the momenta $\OTT P$ and $ \OT P_a$ are primary constraints.

The next step is to determine the variations under $x'^\mu = x^\mu - \epsilon^\mu$. We must bear in mind that the variations of the triads must yield vectors that remain tangent to the fixed time hypersurface. And furthermore the varied $n^\mu = \delta^\mu_0 N^{-1} - \delta^\mu_a N^{-1} N^a$ must be perpendicular to this new hypersurface. The resulting variations are
\beq
\bar \delta N  = N \epsilon^0_{,0} - N N^a \epsilon^0_{,a} + N \epsilon^0_{,0} + N_{,a} \epsilon^a,
\eeq
and
\beq
 \bar \delta N^a =  N^{a}\epsilon^{0}_{,0}
-(N^{2}e^{ab}+N^{a}N^{b})\epsilon^{0}_{,b} +  \epsilon^a_{,0} - N^b \epsilon^a_{,b} + N^{a}_{,0}\epsilon^{0} + N^{a}_{,b}\epsilon^b.
\eeq

To determine the variation of $\tilde T^a_i$ I refer to the variation of the spatial components of the metric. I have
\bea
\bar \delta g_{ab} &=&\bar \delta t^i_a  t^i_b +  t^i_a \bar \delta t^i_b \nonumber \\
&=&  t^i_{a,\mu}\epsilon^\mu  t^i_b +  t^i_a t^i_{b,\mu} \epsilon^\mu + t^i_c N^c \epsilon^\mu_{,a}  t^i_b 
+  t^i_c \epsilon^c_{,a}  t^i_b + t^i_a  t^i_c N^c \epsilon^0_{,b} +t^i_a  t^i_c \epsilon^c_{,b}.
\eea
So I find
\beq
\bar \delta t^i_a = t^i_{a,\mu}\epsilon^\mu + t^i_b N^b \epsilon^0_{,a} +t^i_b \epsilon^b_{,a}
\eeq

Next I calculate $\bar \delta T^a_i$ using
\beq
\bar \delta t^i_a  T^a_j = -  t^i_a  \bar \delta T^a_j,
\eeq
which implies
\bea
\bar \delta T^b_j &=& - \bar \delta t^i_a  T^a_j T^b_i = - \left( t^i_{a,\mu}\epsilon^\mu + t^i_c N^c \epsilon^0_{,a} +t^i_c \epsilon^c_{,a}\right)T^a_j T^b_i \nonumber \\
&=& T^b_{j,\mu} \epsilon^\mu - N^b T^a_j \epsilon^0_{,a} - \epsilon^b_{,a} T^a_j.
\eea
Now to get $\bar \delta \tilde T^a_i$ I need
\beq
\bar \delta t = t \bar \delta t^i_a T^a_i = t \left( t^i_{a,\mu}\epsilon^\mu + t^i_b N^b \epsilon^0_{,a} +t^i_b \epsilon^b_{,a}\right)T^a_i ,
\eeq
which implies
\beq
\bar \delta \tilde T^a_i = \bar \delta t T^a_i + t \bar \delta T^a_i = \tilde T^a_{,\mu} \epsilon^\mu +N^b \epsilon^0_{,b} \tilde T^a_i+ \epsilon^b_{,b} \tilde T^a_i
- N^a \tilde T^c_i \epsilon^0_{,c} - \epsilon^a_{,c} \tilde T^c_i
\eeq

Finally, we also find that
\bea
\bar{\delta} \UT N &=&- \UT N\left( \frac{1}{2} \UT t^i_a \OT T^a_{i,\mu} \epsilon^\mu + \epsilon^a_{,a} + \epsilon^0_{,a} N^a \right) \nonumber \\
&+& \UT N  \epsilon^0_{,0} - \UT N N^a \epsilon^0_{,a} + \left(\frac{1}{2} \UT t^i_a \OT T^a_{i,0} \UT N +  \UT N_{,0}\right) \epsilon^0 + \left(\frac{1}{2} \UT t^i_a \OT T^a_{i,b} \UT N +  \UT N_{,b}\right)\epsilon^b \label{2.32}
\eea

As noted originally in \cite{Pons:1997aa} with regard to Hilbert action, the variations of the lapse and shift are not projectable under the Legendre transformation to phase space due to the dependence on their time derivatives and the unique means of eliminating these terms in spacetime diffeomorphisms  is to require a metric dependence which I rewrite in the form $\OT n^\mu \UT \xi^0$, where
\beq
\OT n^\mu := t n^\mu =\left( \UT N\right)^{-1}\left( \delta^\mu_0 - \delta^\mu_a N^a \right).
\eeq
The general infinitesimal spacetime coordinate variation is therefore
\beq
\epsilon^\mu = t n^\mu \UT \xi^0 + \delta^\mu_a \xi^a.
\eeq

It should be noted here that this requirement results in a loss of the original spacetime diffeomorphism Lie algebra. The most striking change is a forced dependence on the underlying spatial metric, leading to what has become known as the Bergmann Komar group. A detailed history of this development can be found in \cite{Salisbury:2020aa} and \cite{Salisbury:2022ab}.

Taking this required metric dependence into account, the resulting variations are
\bea
\bar \delta N &=& \dot \xi^0 - N^a \xi^0_{,a} + \xi^a N_{,a} = \left( t \UT \xi^0\right)_{,0} - N^a \left(t \UT \xi^0\right)_{,a}+ \xi^a N_{,a}  \nonumber \\
&=&  t_{,0}   \UT \xi^0 + t  \UT \xi^0_{,0}  - N^a t_{,a}\UT \xi^0 -  N^a t \UT \xi^0_{,a}+ \xi^a N_{,a}
\eea
so
\bea
\bar \delta \UT N&=& \bar \delta t^{-1} N +  t^{-1} \bar \delta N \nonumber \\
&=&- t^{-2} \bar \delta t N + t^{-1} \left(t_{,0}   \UT \xi^0 + t  \UT \xi^0_{,0}  - N^a t_{,a}\UT \xi^0 -  N^a t \UT \xi^0_{,a}+ \xi^a N_{,a}\right) \nonumber \\
&=& - t^{-2} \bar \delta t N + t^{-1} \left(t t^i_{a,0}T^a_i  \UT \xi^0 + t  \UT \xi^0_{,0}  - N^a t t^i_{b,a}T^b_i\UT \xi^0 -  N^a t \UT \xi^0_{,a}+ \xi^a N_{,a}\right)
\eea
To continue I need
\bea
\bar \delta t^i_a &=& t^i_{a,\mu}\epsilon^\mu + t^i_b N^b \epsilon^0_{,a} +t^i_b \epsilon^b_{,a} \nonumber \\
&=& N^{-1}t^i_{a,0}\xi^0 - N^{-1}t^i_{a,b}N^b \xi^0 + t^i_{a,b} \xi^b + t^i_b N^b \left(N^{-1} \xi^0\right)_{,a} +t^i_b \left(-N^{-1}N^b \xi^0 + \xi^b\right)_{,a} \nonumber \\
&=& N^{-1}t^i_{a,0}\xi^0 - N^{-1}t^i_{a,b}N^b \xi^0 + t^i_{a,b} \xi^b +t^i_b \left(-N^{-1}N^b_{,a} \xi^0 + \xi^b_{,a}\right) 
\eea
I use this to calculate
\bea
- t^{-2} N \bar \delta t &=&  -t^{-1} N\bar \delta t^i_a T^a_i = -t^{-1} T^a_i \left(t^i_{a,0}\xi^0 - t^i_{a,b}N^b \xi^0 + N t^i_{a,b} \xi^b +t^i_b \left(-N^b_{,a} \xi^0 +N \xi^b_{,a}\right) \right) \nonumber \\
\eea
Combining terms I get
\bea
\bar \delta \UT N&=&  -t^{-1} T^a_i \left(t^i_{a,0}\xi^0 - t^i_{a,b}N^b \xi^0 + N t^i_{a,b} \xi^b +t^i_b \left(-N^b_{,a} \xi^0 +N \xi^b_{,a}\right) \right) \nonumber \\
&+& t^{-1} \left(t t^i_{a,0}T^a_i  \UT \xi^0 + t  \UT \xi^0_{,0}  - N^a t t^i_{b,a}T^b_i\UT \xi^0 -  N^a t \UT \xi^0_{,a}+ \xi^a N_{,a}\right) \nonumber \\
&=& -t^{-1} T^a_i \left( N t^i_{a,b} \xi^b +t^i_b \left(-N^b_{,a} \xi^0 +N \xi^b_{,a}\right) \right) \nonumber \\
&+& t^{-1} \left(  t  \UT \xi^0_{,0}   -  N^a t \UT \xi^0_{,a}+ \xi^a N_{,a}\right) \nonumber \\
&=&- \UT N T^a_i t^i_{a,b} \xi^b +  N^a_{,a} \UT \xi^0 - \UT N \xi^a_{,a} +  \UT \xi^0_{,0} -N^a  \UT \xi^0_{,a} + t^{-1} N_{,a} \xi^a \nonumber \\
&=& N^a_{,a} \UT \xi^0 - \UT N \xi^a_{,a} +  \UT \xi^0_{,0} -N^a  \UT \xi^0_{,a} +  \UT N_{,a} \xi^a 
\eea

Next, I need
\bea
\bar \delta N^a &=& \xi^a_{,0} - N e^{ab} \xi^0_{,b} + N_{,b} e^{ab} \xi^0 + N^a_{,b} \xi^b - N^b \xi^a_{,b} \nonumber \\
&=& \xi^a_{,0} - N e^{ab} \left( t \UT \xi^0\right)_{,b} + \left(t \UT N\right)_{,b} e^{ab}t \UT \xi^0 + N^a_{,b} \xi^b - N^b \xi^a_{,b} \nonumber \\
&=&\xi^a_{,0} -t^2 \UT N e^{ab}  \UT \xi^0_{,b} + t^2 \UT N_{,b} e^{ab} \UT \xi^0 + N^a_{,b} \xi^b - N^b \xi^a_{,b} 
\eea

As a final step I need to consider the variations under $\epsilon^\mu = \delta^\mu_a \xi^a$. These contribute the additional terms to the Noether density
\bea
p^i_a    \tilde T^a_{i,b} \xi^b+ p^i_a\left( \xi^b_{,b} \tilde T^a_i
 - \epsilon^a_{,c} \tilde T^c_i\right)
\eea
After performing an integration by parts, letting $\xi^a \rightarrow 0$ as $x^a \rightarrow \infty$ we obtain the contribution
\bea
&&p^i_a    \tilde T^a_{i,b} \xi^b-\left( p^i_a  \tilde T^a_i \right)_{,b} \xi^b
 +\left( p^i_b \tilde T^a_i\right)_{,a} \xi^b = \left(-p^i_{a,b}  \tilde T^a_i +  p^i_{b,a} \tilde T^a_i +p^i_b \tilde T^a_{i,a}\right) \xi^b \nonumber \\
 &=& 2 D_{[a}p^i_{b]}  \tilde T^a_i \xi^a =: {\cal H}_a.
\eea
Indeed, since $\xi^a$ is an arbitrary spacetime function this delivers an additional vanishing Noether generator of spatial diffeomorphisms.

Substituting the original variations into the Noether charge I obtain
\bea
\mf{C}_\epsilon
&=& p^i_a   \tilde T^a_{i,0} \epsilon^0- {\cal L}_{ADM} \epsilon^0  \nonumber \\
&+& p^i_a    \tilde T^a_{i,b} \epsilon^b+ p^i_a\left(N^b \epsilon^0_{,b} \tilde T^a_i+ \epsilon^b_{,b} \tilde T^a_i
- N^a \tilde T^c_i \epsilon^0_{,c} - \epsilon^a_{,c} \tilde T^c_i\right)+\OTT P \bar \delta \UT N + \OT P_a \bar \delta N^a \nonumber \\
&=& {\cal H}_c \epsilon^0 \nonumber \\
&+& p^i_a    \tilde T^a_{i,b} \epsilon^b+ p^i_a\left(N^b \epsilon^0_{,b} \tilde T^a_i+ \epsilon^b_{,b} \tilde T^a_i
- N^a \tilde T^c_i \epsilon^0_{,c} - \epsilon^a_{,c} \tilde T^c_i\right)+\OTT P \bar \delta \UT N + \OT P_a \bar \delta N^a  \nonumber \\
&=&  \left(\UT N   \left(-{}^3\!R + \frac{1}{4}p^i_a p^i_b e^{ab} - \frac{1}{4}p^i_a T^a_i p^j_b T^b_j\right)  +\frac{1}{2} \left(p^i_a T^a_i  e^{cd}- p^i_a T^d_i e^{ac}\right) t g_{e(c} N^e_{|d)} \right) \epsilon^0 \nonumber \\
&+& p^i_a    \tilde T^a_{i,b} \epsilon^b+ p^i_a\left(N^b \epsilon^0_{,b} \tilde T^a_i+ \epsilon^b_{,b} \tilde T^a_i
- N^a \tilde T^c_i \epsilon^0_{,c} - \epsilon^a_{,c} \tilde T^c_i\right)+\OTT P \bar \delta \UT N + \OT P_a \bar \delta N^a \label{2.17}
\eea

Next collect the terms (\ref{2.17}) involvong $\epsilon^0$ and not the primary constraints. I have
\bea
&& \frac{1}{2} \left(p^i_a T^a_i  e^{cd}- p^i_a T^d_i e^{ac}\right) t g_{e(c} N^e_{|d)}  \epsilon^0 +p^i_a\left(N^b \epsilon^0_{,b} \tilde T^a_i - N^a \tilde T^c_i \epsilon^0_{,c} \right) \nonumber \\
&&=  \frac{1}{2\UT N} \left(p^i_a \tilde T^a_i  e^{cd}- p^i_a \tilde T^d_i e^{ac}\right)   N_{(c|d)}  \UT \xi^0 \nonumber \\
&-&\frac{1}{\UT N}p^i_a\left(-\frac{1}{\UT N} \UT N_{,b} N^b \UT \xi^0_{,b} \tilde T^a_i  + N^b \UT \xi^0_{,b} \tilde T^a_i + \frac{1}{\UT N} \UT N_{,c }N^a \tilde T^c_i \UT \xi^0- N^a \tilde T^c_i \UT \xi^0_{,c} \right) 
\eea
Perform an integration by parts in the first line to get
\bea
&&- \frac{1}{2}\left[\UT N \UT \xi^0 \left(p^i_a \tilde T^a_i  e^{cd}- p^i_a \tilde T^{(d}_i e^{c)a}\right)   \right]_{|d}N_c  \nonumber \\
&&= - \frac{1}{2}\left(\UT N \UT \xi^0 \right)_{|d}\left(p^i_a \tilde T^a_i  N^d- p^i_a \tilde T^{(d}_i N^{a)}\right)  \nonumber \\
&&- \frac{1}{2}\UT N \UT \xi^0  \left(p^i_{a|d} \tilde T^a_i  N^d - p^i_{a|d} \tilde T^{(d}_i N^{a)}\right) 
\eea

In addition I have
\bea
&&p^i_a    \tilde T^a_{i,b} \epsilon^b+ p^i_a\left(\epsilon^b_{,b} \tilde T^a_i
 - \epsilon^a_{,c} \tilde T^c_i\right) \nonumber \\
 &&= p^i_a    \tilde T^a_{i,b} {\UT N}^{-1} N^b \UT \xi^0 -p^i_a \tilde T^a_i \left(-{\UT N}^{-2} \UT N_{,b}N^b{\UT \xi^0+{\UT N}^{-1} N^b_{,b}\UT \xi^0+\UT N}^{-1} N^b\UT \xi^0_{,b}\right) \nonumber \\
 &&+ p^i_a  \tilde T^b_i \left(-{\UT N}^{-2}\UT N_{,b} N^a\UT \xi^0 + {\UT N}^{-1} N^a_{,b}\UT \xi^0 + {\UT N}^{-1} N^a\UT \xi^0_{,b}\right)
\eea

Then it turns out that some amazing cancelations occur, and the resulting Noether charge is
\bea
C_\xi 
&=&  \int d^3x \left[ {\cal H}'_0 \UT \xi^0 + {\cal H}_a  \xi^a\right. \nonumber \\
&+& \left. \OTT P \left(N^a_{,a} \UT \xi^0 - \UT N \xi^a_{,a} +  \UT \xi^0_{,0} -N^a  \UT \xi^0_{,a} +  \UT N_{,a} \xi^a\right) \right. \nonumber \\
&+& \left.  \OT P_a \left(\xi^a_{,0} -t^2 \UT N e^{ab}  \UT \xi^0_{,b} + t^2 \UT N_{,b} e^{ab} \UT \xi^0 + N^a_{,b} \xi^b - N^b \xi^a_{,b} \right) \right] \label{3.30}
\eea
where we have the additional vanishing constraint - due to the arbitrariness in the function $\UT \xi^0$,
\beq
{\cal H}'_0 := -{}^3\!R + \frac{1}{4}p^i_a p^i_b e^{ab} - \frac{1}{4}p^i_a T^a_i p^j_b T^b_j = 0.
\eeq
Similarly, since $\xi^a$ can vary arbitrarily in time, we obtain the constraint
\beq
{\cal H}_a = 0.
\eeq
These results imply, of course, that $C_\xi$ itself vanishes.\footnote{It is likely a surprise to most readers that this procedure for determining what are now known as secondary constraints, following the so-called Bergmann-Dirac procedure, was initiated by L\'eon Rosenfeld in 1930. I and my collaborators believe it would be more accurate to refer to the Rosenfeld-Bergmann-Dirac method. The relation between Bergmann and Dirac is analyzed in detail in \cite{Salisbury:2020aa}, while Rosenfeld's work is discussed in \cite{Salisbury:2017aa} }

\section{Spacetime diffeomorphism-related Noether generator}

I will work out here the requirement to add gauge transformations to the diffeomorphisms in order to attain projectability under the Legendre transformation from configuration-velocity space to phase space.. This challenge arises due to the absence of anti-symmetrized linear combinations of triad time derivatives in the ADM Lagrangian. This is a combination that appears in the Ricci rotation coefficient (See \cite{Pons:2000ac})
\beq
\Omega_0^{ij} = -\tilde T^{a[i}_{,0} \UT t^{j]}_a - N^a_{,b} t^{[i}_a T^{j b} + N^c t^k_c T^{a[i} T^{j]b} t^k_{a,b} + N^c t^{[i}_{c,b} T^{j] b}
\eeq
I undertake the variation of the covector component $\Omega_0^{ij}$ under the infinitesimal diffeomorphism with descriptor $\epsilon^\mu = n^\mu \xi^0 + \delta^\mu_a \xi^a$,
\beq
\bar \delta \Omega_0^{ij} = \Omega_\mu^{ij} \epsilon^\mu_{,0} + \delta \Omega_0^{ij}.
\eeq 
We will not need $\delta \Omega_0^{ij}$ since it is projectible. Thus we have
\beq
\bar \delta \Omega_0^{ij} = \Omega_0^{ij} \left(N^{-1} \xi^0\right)_{,0} +\Omega_a^{ij} \left(-N^{-1}N^a \xi^0 + \xi^a\right)_{,0}+\ldots.
\eeq 
We discover that the unprojectable time derivatives of the lapse and shift appear in this variation. But the good news is that these inadmissible variations can be eliminated by adding gauge rotations with 
\beq
\eta^k = - \epsilon^{kij} \Omega_\mu^{ij} n^\mu \xi^0,
\eeq
with generator
\bea
&&-\int d^3\!x \epsilon^{kij} \Omega_\mu^{ij} n^\mu \xi^0 p_k = -\int d^3\!x \epsilon^{kij} \Omega_\mu^{ij} n^\mu \xi^0 \epsilon^{kmn} p^m_a \tilde T^a_a  \nonumber \\
&&=\int d^3\!x \Omega_\mu^{k[i} \UT t^{j]}_an^\mu \tilde T^a_k \UT \xi^0.
\eea
The additional Ricci rotation coefficient is (from \cite{Pons:2000ac}) the three-dimensional coefficient $\Omega_a^{ij} = \omega_a^{ij}$.

Adding this expression to the first line in (\ref{3.30}) I define the vanishing generator density
\beq
{\cal H}_0 := \left(-{}^3\!R + \frac{1}{4}p^i_a p^i_b e^{ab} - \frac{1}{4}p^i_a T^a_i p^j_b T^b_j + \Omega_\mu^{k[i} \UT t^{j]}_an^\mu \tilde T^a_k\right) = 0.
\eeq

Thus we finally have the full diffeomorphism-related vanishing Noether generator, derived directly from the vanishing Noether charge,
\bea
C_{\xi \eta} 
&=&  \int d^3x \left[ {\cal H}_0 \UT \xi^0 + {\cal H}_a  \xi^a + \eta^k {\cal H}_k\right. \nonumber \\
&+& \left. \OTT P \left(N^a_{,a} \UT \xi^0 - \UT N \xi^a_{,a} +  \UT \xi^0_{,0} -N^a  \UT \xi^0_{,a} +  \UT N_{,a} \xi^a\right) \right. \nonumber \\
&+& \left.  \OT P_a \left(\xi^a_{,0} -t^2 \UT N e^{ab}  \UT \xi^0_{,b} + t^2 \UT N_{,b} e^{ab} \UT \xi^0 + N^a_{,b} \xi^b - N^b \xi^a_{,b} \right) \right] \label{4.7}
\eea

\section{The canonical Hamiltonian}

It  must be stressed that the above diffeomorphism generator differs in an essential manner from the conventional temporal evolution generator. This takes the form
\beq
H = \int d^3\!x \left( \UT N {\cal H}'_0 + N^a {\cal H}_a + \Omega^k {\cal H}_k \right).
\eeq
It evolves initial phase space data in time. The generator $C_{\xi \eta} $, on the other hand, acts on the entire solutions generated by $H$ and transforms them to new physically equivalent solutions that are related through the action of active spacetime diffeomorphisms.  

\section{Extension to the Barbero-Immirzi-Holst model}

The Holst addition to the Lagrangian is
\beq
{\cal L}_H = \frac{1}{4 \gamma} NtE^\mu_I E^\nu_J {}^4\!R^{IJ}_{\mu \nu}
\eeq
 It is introduced  with what has become known as the Barbero-Immirzi parameter $\gamma$. The curvature is expressed in terms of the Ricci rotation coefficients,
 \beq
 {}^4\!R^{IJ}_{\mu \nu} = \partial_\mu \Omega^{IJ}_\nu - \partial_\nu \Omega^{IJ}_\mu + \Omega^{IM}_\mu \Omega_{\nu M}{}^J -\Omega^{IM}_\nu \Omega_{\mu M}{}^J.
 \eeq
 It is of course well known that this Lagrangian vanishes when, as I shall assume, the torsion vanishes. The outcome for my specific use is that the new canonical momentum $p^{\gamma i}_a$is obtained through a canonical transformation of $p^i_a$, i.e.
 \beq
 p^{\gamma i}_a = p^i_a + \frac{1}{2}\gamma^{-1}\epsilon^{ijk} \omega^{jk}_a
 \eeq
 It follows that we need only make this substitution for $p^i_a$ in our Noether generator (\ref{4.7}) to obtain the spacetime diffeomorphism-related symmetry generator in the Barbero-Immirzi-Holst model!
 
 \section{Evolving constants of motion}

I will briefly overview here the manner in which the vanishing diffeomorphism-related generator may be employed to implement the use of intrinsic coordinates, evoking the general method presented in \cite{Pons:2009ab}. There we proposed the use of intrinsic coordinates which must be spacetime scalar phase space functions. I will represent them here as $X^\mu\left(\tilde T^a_i, p^b_j \right)$\footnote{The analogues have long been represented by several authors as $T^\mu$ and they have been denoted as "clock" variables. See for example \cite{Giesel:2018aa}. I would recommend referring to $T^0$ as a clock variable and the $T^a$ rod variables. }. With their aid we can establish gauge conditions which we represent as $\chi^{(1)\mu} = x^\mu - X^\mu = 0$. Recognizing that these must be preserved under time evolution we obtain a second set of gauge conditions 
\beq
0 = \frac{d}{d\,t}\chi^\mu = \delta^\mu_0 -  N^\rho \{X^\mu\,,{\cal H}_\rho\} = \delta^\mu_0  - {\cal A}^\mu_{ \rho}N^\rho =: \chi^{(2) \mu},
\eeq
where
\beq
{\cal A}^\mu_{ \rho} := \left\{X^\mu, {\cal H}_\rho \right\}.
\eeq
In  \cite{Pons:2009ab} we extended a procedure that had been invented by \cite{Dittrich:2007aa} so as to include the lapse and shift as phase space variables. The basic idea is to take linear combinations of the eight first class constraints which I represent here by $\zeta_{(j) \nu} = \left({\cal H}_\mu, \OTT P, \OT P_a\right)$, employing the inverse of ${\cal A}^\mu_{ \rho}$. Representing the new set of the original first class constraints by $\bar \zeta_{(j),\mu}$ we are able to arrange that they satisfy the Poisson brackets with the gauge conditions satisfying 
\beq
\left\{\chi^{(i)\mu} , \bar \zeta_{j,\nu}\right\} = - \delta^i_j \delta^\mu_\nu.
\eeq
Consequently we can solve for the gauge functions $\bar \xi^\mu$ which transform arbitrary solutions of the field equations to those that satisfy the gauge conditions. Of course, in doing so in this case we make use of the generator (\ref{4.7}) with the new linear combinations of constraints $\zeta_{(j) \nu}$. Thus for any phase space function $\Phi$, including the lapse and shift, we can construct the corresponding spacetime  invariant ${\cal I}_\Phi$ through the action of the generator $C_{\bar \xi}$, i.e.
\beq
{\cal I}_\Phi = exp\left( \left\{ - , C_{\bar \xi} \right\}\right) \Phi
\eeq
The validity of this expansion has been demonstrated, for example in \cite{Pons:2009ab}\cite{Salisbury:2022aa}, for several previous models. It will be straightforward to do so for the classical Barbero-Immirzi-Holst theory. A cosmological perturbative approach employing these expansions would be of particular interest. 

\section{Conclusions}

I have presented here a new direct method for obtaining the generator of spacetime diffeomorphism-related phase space transformations through appealing directly to Noether's second theorem. The question that must now be addressed is how one can take these classical symmetries into account in an eventual quantum theory of gravity. Much effort has of course long been devoted to addressing this issue.  Pullin and his collaborators have certainly made significant progress in addressing the associated problem of time \cite{Gambini:2022ah}. Rovelli has long advocated a closely related approach in which a subset of fields serve as clocks. In this regard I and my collaborators are choosing Weyl scalars expressed in terms of phase space variables as both temporal and spatial intrinsic coordinates \cite{Watson:2023aa}. This is accomplished in a manner as advocated in \cite{Pons:2009ab,Pons:2010aa} \cite{Salisbury:2022aa}. But most reassuring is the extension to the full phase space and the corresponding use of intrinsic coordinates that is being pursued in the context of quantum loop cosmology by \cite{Giesel:2018aa} \cite{Giesel:2018ab}\cite{Li:2018aa}\cite{Giesel:2019aa} \cite{Giesel:2019aa} and \cite{Li:2022ac}.

\bibliographystyle{apalike}
\bibliography{qgrav-V19}
\end{document}